\documentstyle[12pt,eqsection]{article}
\newcommand{\be}{\begin{equation}}
\newcommand{\ee}{\end{equation}}
\newcommand{\bea}{\begin{eqnarray}}
\newcommand{\eea}{\end{eqnarray}}
\newcommand{\norsl}{\normalsize\sl}
\newcommand{\norsc}{\normalsize\sc}

\def \ksl {k \kern-.45em{/}}
\def \ppsl {p \kern-.45em{/}}
\def \nsl {n \kern-.45em{/}}

\begin{document}

\begin{titlepage}

\title{ Gauge-independent Thermal $\beta$ Function \\in Yang-Mills Theory 
               \\via the Pinch Technique }
\author{
\norsc  Ken SASAKI\thanks{Permanent address: Dept. of Physics, Yokohama National University, Yokohama 240, Japan. 
e-mail address: sasaki@mafalda.physics.nyu.edu   
or sasaki@ed.ynu.ac.jp}\\
\norsl  Dept. of Physics, New York University\\
\norsl  4 Washington Place, New York, New York 10003, U.S.A.\\}

\date{}
\maketitle
 
\begin{abstract}
{\normalsize It is proposed to use the pinch technique to obtain 
the gauge-independent thermal $\beta$ function $\beta _T$ 
in a hot Yang-Mills gas. 
Calculations of $\beta _T$ are 
performed at one-loop level in four different gauges, 
(i) the background field method with an arbitrary gauge parameter, 
(ii) the Feynman gauge, (iii) the Coulomb gauge, and (iv) 
the temporal axial gauge. When the pinch contributions to 
the gluon self-energy are included, the same result is derived 
for $\beta _T$ in all four cases.
}
\end{abstract}
 

\thispagestyle{empty}
\end{titlepage}
\setcounter{page}{1}
\baselineskip 18pt

\section{Introduction}
\smallskip

It is important for the study of the quark-gluon plasma and/or 
the evolution of the early Universe to fully understand the 
behaviour of the effective coupling constant $\alpha_s (=g^2/{4\pi})$ 
in QCD at high temperature. The running of $\alpha_s$ with the 
temperature $T$ and the external momentum $\kappa$ 
is governed by the thermal $\beta$ function $\beta_T$~\cite{Umezawa}. However, 
the previous calculations of $\beta_T$ have exposed various 
problems: (i) strong vertex dependence---the coupling strongly depends 
on which vertex is chosen to 
renormalize $\alpha_s$~\cite{Niegawa}-\cite{Baier}; 
(ii) severe dependence on the vertex-momentum configuration 
even after a vertex is specified~\cite{Nakkagawa}; 
(iii) the gauge-fixing dependence~\cite{Niegawa}.  
To circumvent these difficulties, it was then proposed to 
use~\cite{Landsman} the Vilkovisky-DeWitt effective 
action~\cite{Vilkovisky}\cite{Rebhan} or to 
use~\cite{ACPS}-\cite{EK} the background field method (BFM) 
for the calculation of $\beta_T$ at one-loop. 
(In Yang-Mills theories the Vilkovisky-DeWitt effective 
action formalism coincides with BFM in the 
background Landau gauge~\cite{Rebhan}.)

First introduced by DeWitt~\cite{Dea}, BFM is a technique for 
quantizing gauge field theories while retaining explicit 
gauge invariance for the background fields. 
Since the Green's functions 
constructed by BFM manifestly maintain gauge invariance, they  
obey the naive QED-like Ward-Takahashi identities and the renormalized 
gauge coupling is defined only through the gluon wave-function 
renormalization constant. As a result, when the static and 
symmetric point is chosen for the renormalization condition of 
the three-gluon vertex, $\beta_T$ is obtained 
in BFM from~\cite{Landsman}\cite{ACPS}-\cite{EK} 
\be
   \beta_T \equiv T\frac{dg(T,\kappa)}{dT}=\frac{g}{2\kappa^2}
              T \frac{d\Pi_T (T,\kappa)}{dT}.
\label{Beta}
\ee
where $\Pi_T (T,\kappa)=\Pi_T(T,k_0=0,\kappa=\vert \vec {\bf k} \vert)$ is 
the transverse function of the gluon self-energy 
$\Pi_{\mu \nu}$ at the static limit. 
Due to the $O(3)$ invariance, the spatial 
part of the gluon self-energy $\Pi_{ij}$ is expressed as follows:
\be
    \Pi_{ij}(k) = \Pi_T (\delta_{ij}-\frac{k_i k_j}{{\vec {\bf k}}^2}) +
                \Pi_L \frac{k_i k_j}{{\vec {\bf k}}^2} 
\ee 
and $\Pi_T$ can be extracted by 
applying the projection operator
\be
           t_{ij}=\frac{1}{2}(\delta_{ij}-\frac{k_i k_j}{{\vec {\bf k}}^2}) 
\ee
to $\Pi_{ij}$. 

The thermal $\beta$ function has been calculated in BFM at one-loop level 
for the cases of the gauge parameter 
$\xi_Q=0$~\cite{Landsman}\cite{Eijck}, 
$\xi_Q=1$~\cite{ACPS} and 
$\xi_Q=$an arbitrary number~\cite{EK}. The results are 
expressed in a form, 
\be
    \beta_T^{BFM}=\frac{g^3 N}{2}\biggl\{
              \frac{7}{16}-\frac{1}{8}(1-\xi_{Q})+
                    \frac{1}{64}(1-\xi_{Q})^2 \biggr\}\frac{T}{\kappa},
\label{BetaBFM}
\ee
where $N$ is the number of colors. 
Contrary to the case of the QCD $\beta$ function at zero temperature, 
$\beta_T^{BFM}$ is dependent on the gauge-parameter $\xi_Q$.
The reason why we have obtained $\xi_Q$-dependent $\beta_T$ in BFM 
is that the contributions to $\beta_T$ come from the finite part of 
the gluon self-energy $\Pi_{\mu\nu}$ and that 
BFM gives $\xi_Q$-dependent finite part for $\Pi_{\mu\nu}$. 
The latter notion had already been known~\cite{Vilkovisky}\cite{Rebhan} 
but was recently brought to light again in a
different context~\cite{Papavass}\cite{SirlinKostas}. 

The purpose of the present paper is to propose the use of 
the pinch technique (PT) to obtain the 
gauge-independent $\beta_T$ in a hot Yang-Mills gas. The preliminary 
results of this paper were given in Ref.\cite{Sasaki}.  
The PT was proposed some time ago by
Cornwall~\cite{rCa} for an algorithm to 
form new gauge-independent proper vertices and new propagators 
with gauge-independent self-energies. 
First it was used to obtain the one-loop 
gauge-independent effective gluon self-energy and vertices in 
QCD~\cite{rCP} and then it has been 
applied to the standard model~\cite{StandardModel}. 
The application of PT to QCD at high temperature was first made by 
Alexanian and Nair~\cite{Nair} to calculate the gap equation for 
the magnetic mass to one-loop order.

In the framework of PT, the one-loop gluon self-energy,  
when the one-loop pinch contributions  
from the vertex and box diagrams are added, 
becomes gauge-independent. In this way we can construct the gauge-independent 
effective gluon self-energy.
As in the case of BFM, the effective gluon self-energy  
constructed by PT obey the naive QED-like Ward-Takahashi identity. 
Thus we can use 
the same Eq.(\ref{Beta}) to calculate $\beta_T$ in the framework of PT. 
More importantly, PT gives the gauge-independent results {\it up to  
the finite terms}, since they are constructed from $S$-matrix.
It was shown recently~\cite{Hiroshima} 
that BFM with the gauge parameter $\xi_Q =1$ reproduces the PT results 
at one-loop order. 
However, for $\xi_Q \not= 1$, this coincidence does not hold any more. 
In fact, BFM gives at one-loop order the gluon self-energy whose 
finite part is $\xi_Q$-dependent. Interestingly enough, 
Papavassiliou~\cite{Papavass} showed that when PT is applied to BFM 
for $\xi_Q \not= 1$ to construct the effective gluon self-energy, 
the gauge-parameter dependence of the finite part disappears and the 
previous $\xi_Q =1$ result (or the universal PT result) is recovered. 
To author's knowledge, there exists, so far, only one approach, i.e. PT, 
which gives the gauge-independent gluon self-energy  
{\it including the finite terms}. 
And indeed these finite terms give contributions to $\beta_T$.
This notion inspires the use of PT for the calculations of $\beta_T$.

The paper is organized as follows. In the next section, we develop 
the general prescription necessary for extracting the pinch contributions 
to the gluon self-energy from 
the one-loop quark-quark scattering amplitude. 
Using this prescription, in Sect.3, we give the {\it complete} expressions of  
the one-loop pinch contributions 
calculated in four different gauges,    
(i) the BFM with an arbitrary gauge parameter, 
(ii) the Feynman gauge, (iii) the Coulomb gauge, and (iv) 
the temporal axial gauge. (The expressions given 
in Ref.\cite{Sasaki} were enough for the calculation of $\beta_T$ but 
not in a complete form.) 
Then we show in detail that when two contributions to $\beta_T$  are added, 
one from the ordinary one-loop gluon self-energy and the other from  
``pinch'' counterpart, we obtain the same $\beta_T$ in the above four 
different gauges. Sect.4 is devoted to the conclusions and discussions.
In addition, 
we present two Appendices. In Appendix A, we give 
one-loop pinch contributions to the gluon self-energy 
from the vertex diagrams of the first kind, of the second kind and 
box diagrams, separately, in the above four different gauges. In Appendix B, 
we list the formulas for thermal one-loop integrals 
necessary for calculating $\beta_T$ in this paper.

\bigskip
\section{Pinch technique}
\smallskip

In this section we explain how to obtain the one-loop pinch 
contributions to 
the gluon self-energy. Let us consider the $S$-matrix element 
$T$ for the elastic quark-quark scattering at one-loop order in the 
Minkowski space, 
assuming that quarks have the same mass $m$. 
Besides the self-energy diagram in Fig.1, 
the vertex diagrams of the first kind and 
the second kind, and the box diagrams, which are shown in 
Fig.2(a), Fig.3(a), and Fig.4(a), respectively, contribute to $T$. 
Such contributions are, in general, gauge-dependent while the sum 
is gauge-independent. Then 
we single out the ``pinch parts'' of the vertex and box diagrams, 
which are depicted in Fig.2(b), Fig.3(b), and Fig.4(b). 
They emerge when a $\gamma^{\mu}$ matrix on the 
quark line is 
contracted with a four-momentum $k_{\mu}$ offered by a gluon propagator or a  
bare three-gluon vertex. Such a term triggers an elementary Ward 
identity of the form
\be
      \ksl = ({\ppsl} + \ksl -m) - ({\ppsl} -m).
\label{Ward}
\ee
The first term removes (pinches out) the internal quark propagator, 
whereas the second term vanishes on shell, or {\it vice versa} . This 
procedure leads to 
contributions to $T$ with one or two less quark propagators and, 
hence, let us call these 
contributions as $T_P$,  ``pinch parts'' of $T$.

Next we extract from $T_P$ the pinch contributions to the gluon self-energy  
$\Pi^{\mu \nu}$. 
First note that the contribution of the gluon self-energy diagram to 
$T$ is written 
in the form 
\be
  T^{(S.E)}=[T^a \gamma ^{\alpha}]D_{\alpha \mu}(k)
       \Pi^{\mu\nu}D^{\nu \beta}(k)[T^a \gamma ^{\beta}],
\ee
where $D(k)$ is a gluon propagator, $T^a$ is a representation matrix  
of $SU(N)$, and $\gamma ^{\alpha}$ and $\gamma ^{\beta}$ are 
$\gamma$ matrices on the external quark lines. 
The pinch contribution $\Pi_P^{\mu \nu}$ to 
$T_P$ should have the same form. 
Thus we must take away $[T^a \gamma ^{\alpha}]D_{\alpha \mu}(k)$ and 
$D^{\nu \beta}(k)[T^a \gamma ^{\beta}]$ from $T_P$.  
For that purpose we use the following identity satisfied by 
the gluon propagator and its inverse:
\bea
   g_{\alpha}^{\beta}&=&D_{\alpha \mu}(k)[D^{-1}]^{\mu \beta}(k)
           =D_{\alpha \mu}(k)[-k^2d^{\mu \beta}] + k_{\alpha}~  {\rm term} 
       \nonumber \\
           &=&D^{-1}_{\alpha \mu }(k)D^{\mu \beta}(k) =
             [-k^2 d_{\alpha \mu}]D^{\mu \beta}(k) + k_{\beta}~ {\rm term},
\label{Identity}
\eea
where
\be
     d^{\mu \nu} = g^{\mu \nu} - \frac{k^{\mu}k^{\nu}}{k^2}.  
\ee 
The $k_{\alpha}$ and  $k_{\beta}$ terms give  null results  when 
they are contracted with $\gamma_{\alpha}$ and of $\gamma_{\beta}$ ,
respectively,  of the external quark lines.

The pinch part of the one-loop vertex diagrams of the first kind depicted 
in Fig.2(b) plus their mirror graphs has a form 
\be
  T_P^{(V_1)}={\cal A}[T^a \gamma ^{\alpha}]D_{\alpha \beta}(k)
      [T^a \gamma ^{\beta}]~,
\ee
where ${\cal A}$ (also ${\cal B}_0$, ${\cal B}_{ij}$, ${\cal C}_0$, and 
${\cal C}_{ij}$ in the equations below) contains a loop integral. 
Using Eq.(\ref{Identity}) we find 
\be
    \gamma^{\alpha}D_{\alpha \beta}(k)\gamma ^{\beta}=
       \gamma^{\alpha}D_{\alpha \mu}(k)[-k^2 d^{\mu \nu}]
     D_{\nu \beta}(k)\gamma^{\beta}
\label{gammagamma}
\ee
Thus  the contributions to  $\Pi^{\mu \nu}$ from the vertex diagrams of 
the first kind are written as 
\be
   \Pi^{\mu \nu (V_1)}_P=[-k^2 d^{\mu \nu}]{\cal A}
\ee
 
The pinch part of the one-loop vertex diagrams of the second kind depicted 
in Fig.3(b) plus their mirror graphs has a form 
\bea
  T_P^{(V_2)}&=&\biggl[T^a \biggl\{ [\gamma^{\kappa}] {\cal B}_0 + 
     \sum_{i,j} [\ppsl_i] p_j^{\kappa} {\cal B}_{ij} \biggr\} \biggr]
   D_{\kappa \beta}(k) [T^a \gamma ^{\beta}]  \nonumber \\
      & &+ (\mu \leftrightarrow  \nu)~,
\eea     
where $(\mu \leftrightarrow  \nu)$ terms are the contributions from 
mirror diagrams, and $p_i$ and $p_j$ are four-momenta appearing 
in the diagrams. 
Using Eq.(\ref{gammagamma}) and 
\be
  [\ppsl_i] p_j^{\kappa}D_{\kappa \beta}(k)=
   [\gamma^{\alpha}]D_{\alpha \mu}(k)[-k^2 d^{\mu \lambda}]p_{i \lambda} 
    p_j^{\nu}D_{\nu \beta}(k)~,
\ee
we obtain  for the contributions to  $\Pi^{\mu \nu}$ from the vertex 
diagram of the second kind
\bea
   \Pi^{\mu \nu (V_2)}_P&=&[-k^2 d^{\mu \nu}]{\cal B}_0 +
        [-k^2 d^{\mu \lambda}]\sum_{i,j} p_{i \lambda} p_j^{\nu} 
      {\cal B}_{ij}  \nonumber \\
      & &+ (\mu \leftrightarrow  \nu)~.
\eea

The pinch part of the one-loop box diagrams depicted 
in Fig.4(b) has a form 
\be
  T_P^{(Box)}=[T^a] \biggl\{ [\gamma^{\alpha}] [\gamma_{\alpha}]{\cal C}_0 + 
     \sum_{i,j} [\ppsl_i][\ppsl_j] {\cal C}_{ij} \biggr\} [T^a].
\ee
Again from Eq.(\ref{Identity}) we see that 
$[\gamma^{\alpha}] [\gamma_{\alpha}]$ and $[\ppsl_i][\ppsl_j]$ are rewritten 
as     
\be
  [\gamma^{\alpha}] [\gamma_{\alpha}] =
     [\gamma^{\alpha}]D_{\alpha \mu}(k)[k^4 d^{\mu \nu}] 
           D_{\nu \beta}(k)[\gamma_{\beta}]              
\ee
\be
    [\ppsl_i][\ppsl_j] =
        [\gamma^{\alpha}]D_{\alpha \mu}(k)[k^4 d^{\mu \lambda}
                d^{\nu \tau} p_{i \lambda} p_{j \tau}]
     D_{\nu \beta}(k)[\gamma_{\beta}]
\ee
and thus we obtain for the contributions to  $\Pi^{\mu \nu}$ from the 
box diagrams
\be
   \Pi^{\mu \nu (Box)}_P=[k^4 d^{\mu \nu}]{\cal C}_0 +
        [k^4 d^{\mu \lambda}d^{\nu \tau}]
        \sum_{i,j} p_{i \lambda} p_{j \tau} 
      {\cal C}_{ij}~.
\ee

It is observed that the prescription developed here is general and can be 
applied to the calculation of the one-loop pinch contributions in any gauge.

\bigskip
\section{Calculation of Thermal $\beta$ Function}
\smallskip

In this section it will be shown that we obtain 
the same $\beta_T$ in the framework of PT 
even when we calculate in four different gauges, 
(i) the background field method with an arbitrary gauge parameter, 
(ii) the Feynman gauge, (iii) the Coulomb gauge, and (iv) 
the temporal axial gauge.  From now on we use the imaginary time 
formalism of thermal field theory. 
Thus the loop integral in 
the Minkowski space is replaced with the following one:
\be
  -i \int \frac{d^4 p}{(2\pi)^4}   \Rightarrow 
        \int dp \equiv \int \frac{d^3 p}{8\pi ^3} T \sum_n,
\ee
where the summation goes over the integer $n$ in $p_0=2\pi inT$. 

\subsection {The Background Field Method}
\smallskip

In the background field method (BFM) with an arbitrary gauge 
parameter $\xi_Q$, the gluon propagator 
$iD_{ab(BFM)}^{\mu \nu}=i\delta _{ab}D_{(BFM)}^{\mu \nu}$ and 
the three-gluon vertex $\widetilde{\Gamma}^{abc}_{\lambda \mu \nu }$ 
with one background gluon field $A^b_{\mu}$  are  given  as 
follows~\cite{Abbott}: 
\be
       D_{(BFM)}^{\mu \nu}=-\frac{1}{k^2} 
            \biggl[ g^{\mu \nu}-(1-\xi_Q)
       \frac{k^\mu k^\nu}{k^2}\biggr],
\label{propagator}
\ee
and
\be
     \widetilde{\Gamma}^{abc}_{\lambda \mu \nu } (p, k, q) = g f^{abc} 
              \biggl[(1-\frac{1}{\xi_Q}) \Gamma^{P}_{\lambda \mu \nu}(p, k, q) 
              + \Gamma^{F}_{\lambda \mu \nu}(p, k, q) \biggr],
\ee
where
\bea
     \Gamma^{P}_{\lambda \mu \nu}(p, k, q) &=& 
               p_{\lambda}g_{\mu \nu} -q_{\nu} g_{\lambda \mu} \nonumber \\
     \Gamma^{F}_{\lambda \mu \nu}(p, k, q) &=&
            2k_{\lambda}g_{\mu \nu}-2k_{\nu}g_{\lambda \mu} 
             -(2p+k)_{\mu}g_{\lambda \nu}, 
\label{Vertex}
\eea
and $f^{bac}$ is the structure constant of the group $SU(N)$. 
In the vertex, $k_\mu$ is taken to be the momentum of the 
background field and each momentum flows inward and, thus, $p+k+q=0$.

The gluon self-energy at one-loop level was calculated in BFM with an 
arbitrary gauge parameter $\xi_Q$~\cite{EHKT} and is given as follows:
\bea
   \Pi^{\mu \nu}_{(BFM)} (k) &=& N g^2 \int dp  \Biggl [ 
    \frac{1}{p^2q^2}(4p^{\mu}p^{\nu}-2p^2 g^{\mu\nu}-k^{\mu}k^{\nu} 
     +4k^2 d^{\mu\nu})  \nonumber \\
   & &\quad -(1-\xi_Q)\frac{k^2}{p^4q^2} 
       \biggl\{ (k^2-q^2)d^{\mu\nu}+\biggl[(d^{\mu\alpha}p_{\alpha}p^{\nu}+
         d^{\nu\beta}p_{\beta}p^{\mu})+(p \leftrightarrow q) \biggr] \biggr\}
               \nonumber \\
  & &\quad +\frac{1}{2}(1-\xi_Q)^2 \frac{k^4}{p^4q^4} 
            d^{\mu\alpha}d^{\nu\beta}p_{\alpha}p_{\beta} \Biggr], 
\label{SelfBFM}
\eea
where it is understood that the loop variables are related by $k + p + q =0$.

The transverse function in the static limit, 
$\Pi_{T}^{(BFG)}(k_0=0, \kappa=\vert \vec {\bf k} \vert)$,  
can be extracted by applying the projection operator $t_{ij}$ 
to $\Pi^{ij}_{(BFM)} (k)$ and we have  
\bea
 \Pi_{T}^{(BFM)}(T, \kappa)&=& 2Ng^{2} \int dp \frac{1}{p^2q^2}
     \biggl[p_0^2 +2\vec {\bf k}^2- \frac{(\vec {\bf k} \cdot \vec {\bf p})^2}
         {\vec {\bf k}^2} \biggr]  \nonumber  \\
   & &+Ng^{2} (1-\xi_{Q})  \int dp \Biggl\{
      \frac{\vec {\bf k}^2}{p^2q^2}+\frac{2}{p^4q^2}\biggl[ 
    -\vec {\bf k}^2(\vec {\bf k} \cdot \vec {\bf p}) +
    \vec {\bf k}^2 \vec {\bf p}^2-
     (\vec {\bf k} \cdot \vec {\bf p})^2  \biggr] \Biggr\}
 \nonumber \\
   & &+\frac{N}{4}g^2(1-\xi_{Q})^2 \vec {\bf k}^2 \int dp 
          \frac { \vec {\bf k}^2 \vec {\bf p}^2-
     (\vec {\bf k} \cdot \vec {\bf p})^2}{p^4 q^4}.
\label{PiTBFM}
\eea
After the $p_0$ summation and the angular integration we obtain 
\bea
 \Pi_{T}^{(BFM)}(T, \kappa)&=&
    \frac{Ng^2}{4\pi^2}\int_{0}^{\infty} dp~p~n(p)
        \Biggl[-2+\biggl(\frac{2p}{\kappa}+ \frac{7\kappa}{2p} \biggr)
       ~{\rm ln}  \Big\vert \frac{2p + \kappa}{2p - \kappa} \Big\vert 
      \Biggr]  \nonumber \\
& &-Ng^2 (1-\xi_{Q})  \frac{\kappa}{4\pi^2}
        \int_{0}^{\infty} dp ~p~n(p)
 \biggl\{ - \frac{4p \kappa}{(2p+\kappa)(2p-\kappa)} + 
   {\rm ln}  \Big\vert \frac{2p + \kappa}{2p - \kappa} \Big\vert \biggr\}
   \nonumber \\
& &+\frac{N}{4}g^2(1-\xi_{Q})^2  \nonumber \\
 & & \qquad \times \frac{\kappa}{4\pi^2}
        \int_{0}^{\infty} dp ~p~n(p) \biggl\{ -
   \frac{2\kappa^2}{(2p+\kappa)(2p-\kappa)}+\frac{\kappa}{2p}~{\rm ln} 
        \Big\vert \frac{2p + \kappa}{2p - \kappa} \Big\vert 
      \biggr\}   \nonumber  \\
\eea
where, in the r.h.s., $p\equiv \vert \vec {\bf p} \vert$, 
distribution function, and 
we have used formulas given in Appendix B for the thermal one-loop integrals.  
Also listed in Appendix B are the formulas 
useful for the $\kappa<<T$ expansion. 
In the limit $\kappa<<T$ we get 
\be
     \Pi_{T}^{(BFM)}(T, \kappa) \approx Ng^{2}{\kappa}T \biggl\{
              \frac{7}{16}-\frac{1}{8}(1-\xi_{Q})+
                    \frac{1}{64}(1-\xi_{Q})^2 \biggr\} + {\cal O} (\kappa^2).
\label{BFM}
\ee
Using this expression for $\Pi_{T}$ in Eq.(\ref{Beta}), Elmfors and Kobes 
obtained Eq.(\ref{BetaBFM}) for $\beta_T^{BFM}$ 
which is indeed gauge-parameter $\xi_Q$ dependent~\cite{EK}.

Now we evaluate the pinch contributions to $\Pi_{T}$. 
We consider the quark-quark scattering at one-loop order,  
using the gluon propagator and the three-gluon 
vertex given in Eqs.(\ref{propagator})-(\ref{Vertex}). 
We pinch out the internal quark propagators and obtain the pinch parts of 
the scattering amplitude $T_P$.
Since the inverse of the propagator $D_{(BFM)}^{\mu \nu}(k)$ is given by
\be
     {[}D^{-1}_{(BFM)}{]}^{\mu \nu}(k)
        = -k^2\biggl[ g^{\mu \nu}-
       (1-\frac{1}{\xi_Q}) \frac{k^\mu k^\nu}{k^2}   \biggr], 
\ee
$D_{(BFM)}$ and $D^{-1}_{(BFM)}$ satisfy the relations in Eq.(\ref{Identity})
\bea
       D^{(BFM)}_{\alpha \mu}(k)[D^{-1}_{(BFM)}]^{\mu \beta}(k)&=&
    D^{(BFM)}_{\alpha \mu}(k)[-k^2d^{\mu \beta}] + 
       \frac{k_{\alpha}k^{\beta}}{k^2}     \nonumber  \\
       D^{-1}_{(BFM)\alpha \mu }(k)D^{\mu \beta}_{(BFM)}(k) 
          &=& [-k^2 d_{\alpha \mu}]D^{\mu \beta}_{(BFM)}(k) 
          +  \frac{k_{\alpha}k^{\beta}}{k^2} .
\eea
Thus we can follow the prescription explained in Sect.2 and 
extract the pinch contributions to the gluon self-energy 
from $T_P$. The sum is expressed as   
\bea
   \Pi^{\mu \nu}_{P(BFM)} (k) &=& N g^2 \biggl [ (1-\xi_{Q}) k^2 
              d^{\mu \nu} \int dp 
             \frac{-2kp}{p^4 q^2}   \nonumber  \\
    & &+\frac{1}{2} (1-\xi_{Q})^2 k^4  d^{\mu \alpha} d^{\nu \beta} \int dp
     \frac{-p_{\alpha}p_{\beta}}{p^4 q^4} \biggr ], 
\label{SelfPBFM}
\eea
which was first obtained by Papavassiliou~\cite{Papavass}. 
For completeness the one-loop pinch contributions in BFM from 
the vertex diagrams of the first kind [Fig.2(b) and its mirror graph], 
the vertex diagrams of the second kind [Fig.3(b) and its mirror graph] 
and the box-diagrams [Fig.4(b)] are separately given in Appendix A.

Applying the projection operator $t_{ij}$ to the spatial part of 
$\Pi^{\mu \nu}_{P(BFG)}(k_0=0, \kappa=\vert \vec {\bf k} \vert)$,  
we obtain 
\bea
     \Pi_{T}^{P(BFM)}(T,\kappa)& &=N g^2 (1-\xi_{Q}) {\kappa}^2 
 \int dp \frac{2{\vec {\bf k}}\cdot {\vec {\bf p}}}
            {p^4 q^2}    \nonumber   \\
        & & - \frac{N}{4}g^2 (1-\xi_{Q})^2 {\kappa}^2 
           \int dp 
      \frac{\vec {\bf k}^2 \vec {\bf p}^2 - ({\vec {\bf k}}
    \cdot {\vec {\bf p}})^2}
           {p^4 q^4}~.
\label{PiTPBFM}
\eea
Clearly, the $(1-\xi_{Q})^2$ terms of $\Pi_{T}^{(BFM)}$ in 
Eq.(\ref{PiTBFM}) and of  $\Pi_{T}^{P(BFM)}$ in Eq.(\ref{PiTPBFM}) 
are the same but have an opposite sign, and so when they are combined 
they are cancelled out. Also we can see 
the $(1-\xi_{Q})$ terms of $\Pi_{T}^{(BFM)}$  
and of  $\Pi_{T}^{P(BFM)}$ cancel when combined, due to the identity 
\be
 \int dp \frac{ \vec {\bf k}^2 \vec {\bf p}^2-
     (\vec {\bf k} \cdot \vec {\bf p})^2}{p^4 q^2} =
    - \frac{1}{2} \vec {\bf k}^2 \int dp \frac{1}{p^2 q^2}.
\ee  
Thus we find the sum 
\bea 
\Pi_{T}(T,\kappa)& &\equiv \Pi_{T}^{(BFM)}(T,\kappa)+
    \Pi_{T}^{P(BFM)}(T,\kappa)  \nonumber  \\
   & &= \frac{Ng^2}{4\pi^2}\int_{0}^{\infty} dp~p~n(p)
        \Biggl[-2+\biggl(\frac{2p}{\kappa}+ \frac{7\kappa}{2p} \biggr)
       ~{\rm ln}  \Big\vert \frac{2p + \kappa}{2p - \kappa} \Big\vert 
      \Biggr]       
\label{InvPi}
\eea
is  gauge-parameter $\xi_Q$ independent. In the limit $\kappa << T$ 
\be
   \Pi_{T}(T,\kappa) \approx Ng^{2}{\kappa}T \frac{7}{16} 
      + {\cal O} (\kappa^2)
\label{InvPiApprox}
\ee
and this gives a $\xi_Q$-independent thermal $\beta$ function 
\be
      \beta_T=g^3 N\frac{7}{32}\frac{T}{\kappa}.
\label{InvBeta} 
\ee
Actually we will see below that the expression is independent of the 
choice of gauge-fixing. 
Note that the result coincides with $\beta_T^{BFM}$ in 
Eq.(\ref{BetaBFM}) with $\xi_Q=1$~\cite{ACPS}.

It should be remarked that not only the sum of 
$\Pi_{T}^{(BFM)}$ and $\Pi_{T}^{P(BFM)}$, but also 
the sum of the one-loop gluon self-energy $\Pi^{\mu \nu}_{(BFM)} (k)$ 
and the corresponding pinch contribution $\Pi^{\mu \nu}_{P(BFM)} (k)$ 
becomes $\xi_Q$-independent~\cite{Papavass}. Indeed 
the $(1-\xi_Q)$ term of $\Pi^{\mu \nu}_{(BFM)} (k)$ in Eq.(\ref{SelfBFM})
can be rewritten as 
\be 
    N g^2  (1-\xi_{Q}) k^2  d^{\mu \nu} \int dp 
             \frac{2kp}{p^4 q^2}   
\ee
with the help of the identity
\be
   \int dp \frac{4p_{\mu}p_{\nu}}{p^4q^2}= \int dp \frac{g_{\mu\nu}}{p^2q^2}
     - \int dp \frac{2k_{\mu}p_{\nu}}{p^4q^2},
\ee
where $p+q+k=0$ is understood and 
\be
       d_{\mu\alpha} \int dp \frac{p^{\alpha}}{p^4q^2}=0.
\ee
Thus both the $(1-\xi_Q)$ and $(1-\xi_Q)^2$ terms of 
$\Pi^{\mu \nu}_{(BFM)} (k)$ in Eq.(\ref{SelfBFM}) cancel out 
with the corresponding ones in 
$\Pi^{\mu \nu}_{P(BFM)} (k)$ in Eq.(\ref{SelfPBFM}) and 
there remains the $\xi_Q$ independent 
(actually it is independent of the choice of gauge-fixing~\cite{Passera}) 
expression for the effective gluon self-energy,
\be
 \widehat{\Pi}^{\mu \nu} (k) = N g^2 \int dp   
    \frac{1}{p^2q^2}(4p^{\mu}p^{\nu}-2p^2 g^{\mu\nu}-k^{\mu}k^{\nu} 
     +4k^2 d^{\mu\nu}). 
\label{InvSelf}
\ee
Of course we can get the result of Eq.(\ref{InvPi}) directly  by 
applying $t_{ij}$ to the above $\widehat{\Pi}^{\mu \nu} (k)$ and by 
integration.

It is amusing to note that in BFM with the choice of $\xi_Q=1$ there 
is no pinch contribution to the gluon self-energy at one-loop order. 
This is due to the fact 
with $\xi_Q=1$ the longitudinal term $k_{\mu}k_{\nu}$ of 
the gauge boson propagator disappears and the three-gluon 
vertex $\widetilde{\Gamma}^{abc}_{\lambda \mu \nu }$ is made up of 
only $\Gamma^F_{\lambda \mu \nu }$ and, hence, there appears no four-momentum 
which triggers the Ward identity of Eq.(\ref{Ward}) and pinches out a 
quark propagator. Therefore 
when we work with the choice of $\xi_Q=1$ in BFM and calculate the gluon 
self-energy diagrams, we directly obtain the the gauge-independent 
expression  $\widehat{\Pi}^{\mu \nu} (k)$ in 
Eq.(\ref{InvSelf})~\cite{Hiroshima}.

\bigskip
\noindent
\subsection{\it The Feynman Gauge}

In the Feynman gauge (FG) (the covariant gauge with $\xi = 1$) 
the gluon propagator, 
$iD_{ab(FG)}^{\mu \nu}=i\delta _{ab}D_{(FG)}^{\mu \nu}$,  
has a very simple form
\be
   D_{(FG)}^{\mu \nu}(k)=\frac{-1}{k^2}
                 g^{\mu \nu},
\ee
and the three-gluon vertex is expressed as
\be
\Gamma^{abc}_{\lambda \mu \nu } (p, k, q) = g f^{bac} 
              \biggl[ \Gamma^{P}_{\lambda \mu \nu}(p, k, q) 
              + \Gamma^{F}_{\lambda \mu \nu}(p, k, q) \biggr]
\label{VertexFG}
\ee
where $\Gamma^{P}_{\lambda \mu \nu}(p, k, q)$ and 
$\Gamma^{F}_{\lambda \mu \nu}(p, k, q)$ are given in Eq.(\ref{Vertex}).
The expression of the one-loop gluon self-energy in FG is well known: 
\be
\Pi^{\mu\nu}_{(FG)}(k)=N g^2 \int dp \frac{1}{p^2 q^2} 
   \biggl[4p^{\mu}p^{\nu}-2p^2 g^{\mu\nu}-k^{\mu}k^{\nu} 
              +2k^2 d^{\mu\nu} \biggr].
\ee

Since the inverse of the gluon propagator is given by 
\be
      {[}D^{-1}_{(FG)}{]}^{\mu \nu}(k)
        = -k^2 g^{\mu \nu},
\ee
$D_{(FG)}$ and its inverse satisfy 
\bea
       D^{(FG)}_{\alpha \mu}(k)[D^{-1}_{(FG)}]^{\mu \beta}(k)&=&
    D^{(FG)}_{\alpha \mu}(k)[-k^2d^{\mu \beta}] + 
       \frac{k_{\alpha}k^{\beta}}{k^2}    \nonumber  \\
       D^{-1}_{(FG)\alpha \mu }(k)D^{\mu \beta}_{(FG)}(k) 
          &=& [-k^2 d_{\alpha \mu}]D^{\mu \beta}_{(FG)}(k) 
          +  \frac{k_{\alpha}k^{\beta}}{k^2}  .  
\eea
Then we follow the prescription explained in Sect.2 and we obtain the 
pinch contributions to the gluon self-energy.
Since the gluon propagator in FG does not have a longitudinal  
$k^{\mu}k^{\nu}$ term, the only contribution is coming from 
the vertex diagram of the second kind with the three-gluon 
vertex $\Gamma^{P}$ (and its mirror graph)~\cite{rCP}, and it is given by 
\be
  \Pi^{\mu \nu}_{P(FG)}(k) =2N g^2 k^2 d^{\mu \nu}\int dp  
             \frac{1}{p^2 q^2} .
\ee
Adding $\Pi^{\mu \nu}_{(FG)}$ and $\Pi^{\mu \nu}_{P(FG)}$, we find the sum
coincides with the 
gauge independent $\widehat{\Pi}^{\mu \nu} (k)$ in Eq.(\ref{InvSelf}). 
Thus we obtain the same $\Pi_{T}$ in Eq.(\ref{InvPiApprox}) and the 
same $\beta_T$ 
in Eq.(\ref{InvBeta}) while we work in FG.

\bigskip
\noindent
\subsection{\it The Coulomb Gauge}

In fact it is rather anticipated that 
once we use PT for the one-loop calculation of the thermal 
$\beta$ function, we obtain the $\xi_Q$-independent $\beta_T$ in BFM 
which coincides with the result in FG.
However, it is less trivial whether we may reach the same result 
for  $\beta_T$ when we calculate in noncovariant gauges such as 
the Coulomb gauge and the temporal axial gauge.  We show 
in this and the following subsections that we indeed obtain the same 
$\beta_T$ in the above two noncovariant gauges when we use PT.

Given a unit vector $n^{\mu}=(1,0,0,0)$, 
the gluon propagator in the Coulomb gauge (CG),
$iD_{ab(CG)}^{\mu \nu}=i\delta _{ab}D_{(CG)}^{\mu \nu}$,  
and its inverse are expressed as 
\bea
    D_{(CG)}^{\mu \nu}(k)&=&-\frac{1}{k^2}\biggl[g^{\mu \nu}+
     \biggl(1-\xi_C \frac{k^2}{{\vec {\bf k}}^2} \biggr) 
                \frac{k^{\mu}k^{\nu}}{{\vec {\bf k}}^2} -
      \frac{k_0}{{\vec {\bf k}}^2}(k^{\mu}n^{\nu}+n^{\mu}k^{\nu})\biggr] 
   \\
   {[}D^{-1}_{(CG)}{]}^{\mu \nu}(k)&=&-k^2\biggl[g^{\mu \nu}-
        \frac{k^{\mu}k^{\nu}}{k^2}\biggr]+\frac{1}{\xi_C}
        \biggl[k^{\mu}k^{\nu}-k_0(k^{\mu}n^{\nu}+n^{\mu}k^{\nu})+
                         k_0^2 n^{\mu}n^{\nu}\biggr]~.  \nonumber 
\eea
where $\xi_C$ is the gauge parameter of CG. 
The three-gluon vertex is the same as in FG, that is, 
$\Gamma^{abc}_{\lambda \mu \nu } (p, k, q)$ in Eq.(\ref{VertexFG}). 
It is noted that although in the limit $\xi_C=0$, 
$D_{(CG)}^{\mu \nu}(k)$ reduces to the well-known form~\cite{Ali}
\be
  D_{(CG)}^{00}=\frac{1}{{\vec {\bf k}}^2}, \quad 
  D_{(CG)}^{0i}=0, \quad   
  D_{(CG)}^{ij}=\frac{1}{k^2}\biggl(\delta^{ij}-
       \frac{k^i k^j}{{\vec {\bf k}}^2}\biggr)~,
\label{CGpropagatorlimit} 
\ee
its inverse does not exist in this limit. 
In the framework of PT, we need to use the identities in Eq.(\ref{Identity}), 
satisfied by 
the gluon propagator and its inverse, to extract from $T_P$ 
the pinch contributions to 
the gluon self-energy. Thus in principle we must work 
with a non-zero  $\xi_C$. 
Then at one-loop level there appear 
$\xi_C$-dependent terms in the gluon self-energy. 
We can  show~\cite{Passera}, however,  
that the one-loop pinch contributions are also 
$\xi_C$-dependent and these $\xi_C$-dependent parts exactly cancel 
against the $\xi_C$-dependent terms  in the self-energy. 
For our purpose of calculating the thermal $\beta$ function, therefore, 
it is enough to know the information 
on the $\xi_C$-independent part of both the 
self-energy and the pinch contributions. The $\xi_C$-independent part of the 
one-loop gluon self-energy was calculated in Ref.\cite{HKT} 
using the gluon propagator in the 
$\xi_C=0$ limit given in Eq.(\ref{CGpropagatorlimit}).

The transverse function $\Pi_T$ is related to 
the self-energy as 
\be
   \Pi_T=t_{ij}\Pi_{ij}=\frac{1}{2}\biggl[\Pi_{ii}-
           \frac{1}{\vec {\bf k}^2}k_i \Pi_{ij}k_j \biggr].
\ee
Since  $k_i \Pi_{ij}k_j=0$ in the static limit $k_0=0$, we have 
$\Pi_T(T,\kappa)=\frac{1}{2}\Pi_{ii}(k_0=0, \kappa)$. 
The $\xi_C$-independent part of $\Pi^{(CG)}_{ii}(k)$   
is given in Eq.(4.12) of Ref.\cite{HKT} as 
\bea
\Pi_{ii}^{(CG)}(k) &=&\frac{N}{2}g^2 \int dp
  \Biggl\{ \frac{8}{p^2}-\frac{6}{{\vec {\bf p}}^2}  \nonumber \\
 & &\quad + \frac{1}{p^2q^2}\biggl[\vert{\vec {\bf p}}- {\vec {\bf q}}\vert^2 
    (1+c^2)+8({\vec {\bf p}}^2+{\vec {\bf q}}^2)(1-c^2) \biggr] 
  \nonumber \\
  & &\quad +\biggl[2\frac{(k_0-p_0)^2}{p^2{\vec {\bf q}}^2} +
   (p \leftrightarrow  q)  \biggr] +
   \frac{\vert{\vec {\bf p}}- {\vec {\bf q}}\vert^2}{{\vec {\bf p}}^2
   {\vec {\bf q}}^2}+\frac{2{\vec {\bf p}} \cdot {\vec {\bf q}}}
    {{\vec {\bf p}}^2 {\vec {\bf q}}^2} \Biggr\}
\eea
Using the formulas given in Appendix B, 
we can calculate the static limit of 
$\Pi_{ii}^{CG}$ and we obtain 
\bea
     \Pi_{T}^{(CG)}(T,\kappa)&=&\frac{1}{2}\Pi_{ii}^{(CG)}(k_0=0,\kappa) 
     \nonumber  \\
 &=& \frac{Ng^2}{4\pi^2} 
      \int_{0}^{\infty} dp ~p~n(p) 
   \Biggl\{-\frac{3}{2}+\frac{5\kappa^2}{4p^2} + 
   \biggl[-\frac{p}{4\kappa}-\frac{11\kappa}{8p}+\frac{\kappa^3}{2p^3}
 +\frac{\kappa^5}{8p^5}\biggr]~{\rm ln} 
        \Big\vert \frac{p + \kappa}{p - \kappa} \Big\vert 
 \nonumber  \\
  & &\qquad  \qquad \qquad +\biggl[\frac{2p}{\kappa}+\frac{5\kappa}{2p}
 -\frac{\kappa^3}{2p^3}-\frac{\kappa^5}{16p^5} \biggr] 
 ~{\rm ln} 
        \Big\vert \frac{2p + \kappa}{2p - \kappa} \Big\vert
  \Biggr\}
\label{CG}
\eea
In the limit $\kappa << T$, this gives $\Pi_{T}^{(CG)}(T,\kappa) 
\approx  Ng^{2}{\kappa}T\frac{9}{64} + {\cal O} (\kappa^2)$.

Now we calculate the pinch contributions to the gluon self-energy in CG.
Since the gluon propagator and its inverse satisfy the relations in 
Eq.(\ref{Identity}), that is, 
\bea
       D^{(CG)}_{\alpha \mu}(k)[D^{-1}_{(CG)}]^{\mu \beta}(k)&=&
    D^{(CG)}_{\alpha \mu}(k)[-k^2d^{\mu \beta}] + 
       \frac{k_{\alpha}}{{\vec {\bf k}}^2}(k_0 n^{\beta}- k^{\beta})  
\nonumber  \\
       D^{-1}_{(CG)\alpha \mu }(k)D^{\mu \beta}_{(CG)}(k) 
          &=& [-k^2 d_{\alpha \mu}]D^{\mu \beta}_{(CG)}(k)+ 
         (k_0 n_{\alpha}- k_{\alpha})\frac{k^{\beta}}{{\vec {\bf k}}^2}~, 
\eea
again we can follow the prescription explained in Sec.2 to extract 
the one-loop pinch contributions.
The individual contributions in CG from the vertex diagrams of the first kind, 
of the second kind and the box diagrams are presented in 
Appendix A.  In total we obtain for the $\xi_C$-independent part
\bea
  \Pi^{\mu \nu}_{P(CG)}(k) &=& N g^2  k^2 d^{\mu \nu} \int dp 
     \frac{1}{p^2 q^2 {\vec {\bf p}}^2} (k^2 - q^2 -
 4 {\vec {\bf k}}\cdot {\vec {\bf p}})   \nonumber \\
 &+&\frac{N}{2}g^2  k^2 d^{\mu \alpha} d^{\nu \beta}\int dp 
\frac{1}{p^2 q^2 {\vec {\bf p}}^2 {\vec {\bf q}}^2}
\biggl\{p_{\alpha}p_{\beta}(k^2 - 4{\vec {\bf q}}^2 +2{\vec {\bf k}}^2)
           \nonumber \\ 
& &\quad  +(p_{\alpha}n_{\beta}+n_{\alpha}p_{\beta})
      \bigl[p^2q_0-q^2p_0-2{\vec {\bf p}}\cdot {\vec {\bf q}}(p_0-q_0)\bigr]
     +n_{\alpha}n_{\beta} 4p_0q_0 (p q) \biggr\} \nonumber \\
&+& \frac{N}{2} g^2 \Biggl[ d^{\mu \alpha} \int dp 
  \biggl\{ p_{\alpha}k^{\nu} \biggl[ \frac{1}{q^2 \vec {\bf p}^2} - 
      \frac{1}{p^2 \vec {\bf q}^2} + 
   \bigl( \frac{1}{q^2}-\frac{1}{p^2} \bigr) 
   \frac{{\vec {\bf p}}\cdot {\vec {\bf q}} }
   {\vec {\bf p}^2 \vec {\bf q}^2}  \biggr]  \nonumber \\ 
& & \qquad  + n_{\alpha}k^{\nu}\biggl[ 
  - \frac{q_0}{p^2 {\vec {\bf q}}^2} -\frac{p_0}{q^2 {\vec {\bf p}}^2}
     +   \bigl(\frac{q_0}{q^2}+\frac{p_0}{p^2} \bigr) 
      \frac{{\vec {\bf p}}\cdot {\vec {\bf q}}}
       {  {\vec {\bf p}}^2 {\vec {\bf q}}^2}  \biggr] \biggr\}
         + (\mu \leftrightarrow \nu) \Biggr].
\label{PinchCG}
\eea

It was pointed out~\cite{HKT} that the gluon self-energy 
caluculated in CG does not satisfy the transversality relation, i.e.,
$\Pi^{\mu \nu}_{(CG)}k_{\nu} \ne 0$. Now we can see that the pinch 
contribution  
$\Pi^{\mu \nu}_{P(CG)}$ does not satisfy the transversality relation either. 
In fact we get 
\bea
\Pi^{\mu \nu}_{P(CG)}k_{\nu}&=& \frac{N}{2} g^2 k^2
    d^{\mu \alpha} \int dp 
    \biggl\{ p_{\alpha}\biggl[ \frac{1}{q^2 \vec {\bf p}^2} - 
      \frac{1}{p^2 \vec {\bf q}^2} + 
   \bigl( \frac{1}{q^2}-\frac{1}{p^2} \bigr) 
   \frac{{\vec {\bf p}}\cdot {\vec {\bf q}} }
   {\vec {\bf p}^2 \vec {\bf q}^2}  \biggr]  \nonumber \\ 
& & \qquad  + n_{\alpha}\biggl[ 
  - \frac{q_0}{p^2 {\vec {\bf q}}^2} -\frac{p_0}{q^2 {\vec {\bf p}}^2}
     +   \bigl(\frac{q_0}{q^2}+\frac{p_0}{p^2} \bigr) 
      \frac{{\vec {\bf p}}\cdot {\vec {\bf q}}}
       {  {\vec {\bf p}}^2 {\vec {\bf q}}^2}  \biggr] \biggr\}~.
\eea  
It can be shown that this non-transverse part of pinch contribution 
exactly cancels against the non-transverse part of
$\Pi^{\mu \nu}_{(CG)}k_{\nu}$ and that the sum of 
$\Pi^{\mu \nu}_{(CG)}$ and $\Pi^{\mu \nu}_{P(CG)}$ indeed satisfies the 
transversality relation~\cite{Passera}.

The function  $\Pi_{T}^{P(CG)}(T,\kappa)$ is obtained by 
applying the projection operator $t_{ij}$ to the spatial part of  
$\Pi^{ij}_{P(CG)}(k_0=0,\kappa)$. 
Since $d^{i\alpha}n_{\alpha}=0$ in the static limit and $t_{ij}k^j=0$, 
the terms proportional to $(p_{\alpha}n_{\beta}+n_{\alpha}p_{\beta})$, 
$n_{\alpha}n_{\beta}$, $d^{\mu \alpha}p_{\alpha}k^{\nu}$ and 
$d^{\mu \alpha}n_{\alpha}k^{\nu}$ in Eq.(\ref{PinchCG}) do not 
contribute to $\Pi_{T}^{P(CG)}$. The result is 
\bea
     \Pi_{T}^{P(CG)}(T,\kappa)&=&-N g^2 {\kappa}^2 
 \int dp \biggl\{ \frac{1}{q^2 \vec {\bf p}^2}+ 
   \frac{2{\vec {\bf k}}\cdot {\vec {\bf p}}}{p^2 q^2 \vec {\bf p}^2} \biggr\} 
    \nonumber   \\
   & &- \frac{N}{4}g^2{\kappa}^2 \int dp  
    \biggl[1 - \frac{({\vec {\bf k}}\cdot {\vec {\bf p}})^2}
         {\vec {\bf k}^2 \vec {\bf p}^2}\biggr] 
   \biggl\{\frac{\vec {\bf k}^2}{p^2 q^2 \vec {\bf q}^2} 
            - \frac{4}{p^2 q^2} \biggr\}~. 
\eea
Since the terms proportional to $n$ and $k$ do not contribute, 
the quickiest way to arrive at the above expression of $\Pi_{T}^{P(CG)}$  
is that we consider the 
quark-quark scattering amplitude at one-loop order and discard the terms which 
are proportional to $\nsl$ and $\ksl$ from the beginning. In fact, this 
simplified method was used in Ref.\cite{Sasaki} to obtain the 
pinch contribution $\Pi_{T}^{P}$ in both the Coulomb gauge and the 
temporal axial gauge.

After the $p_0$ summation and the angular integration for $\Pi_{T}^{P(CG)}$ 
(see the formulas listed in Appendix B), we have 
\bea
  \Pi_{T}^{P(CG)}(T,\kappa)&=& \frac{Ng^2}{4\pi^2} 
      \int_{0}^{\infty} dp ~p~n(p) 
   \Biggl\{-\frac{1}{2}-\frac{5\kappa^2}{4p^2} + 
   \biggl[\frac{p}{4\kappa}+\frac{11\kappa}{8p}-\frac{\kappa^3}{2p^3}
 -\frac{\kappa^5}{8p^5}\biggr]~{\rm ln} 
        \Big\vert \frac{p + \kappa}{p - \kappa} \Big\vert
  \nonumber  \\  
  & &\qquad  \qquad \qquad +\biggl[\frac{\kappa}{p}
 +\frac{\kappa^3}{2p^3}+\frac{\kappa^5}{16p^5} \biggr] 
 ~{\rm ln} 
        \Big\vert \frac{2p + \kappa}{2p - \kappa} \Big\vert
  \Biggr\}.
\label{PCG}
\eea
In the limit $\kappa <<T$, this gives 
$ \Pi_{T}^{P(CG)}(T,\kappa) \approx 
           Ng^{2}{\kappa}T\frac{19}{64} + {\cal O} (\kappa^2)$. 
Adding the two contributions, $\Pi_{T}^{(CG)}$ and $\Pi_{T}^{P(CG)}$, 
we find that the sum is equal to $\Pi_{T}$ in Eq.(\ref{InvPi}) and, thus, 
we obtain the same $\beta_T$ in Eq.(\ref{InvBeta}).  

\bigskip
\noindent
\subsection{\it The Temporal Axial Gauge}

The gluon propagator in the temporal axial gauge (TAG),  
$iD_{ab(TAG)}^{\mu \nu}=i\delta _{ab}D_{(TAG)}^{\mu \nu}$, and its 
inverse are defined by
\bea
    D_{(TAG)}^{\mu \nu}(k)&=&-\frac{1}{k^2}\biggl[g^{\mu \nu}+
     (1+\xi_A k^2) 
                \frac{k^{\mu}k^{\nu}}{k_0^2} -
      \frac{1}{k_0}(k^{\mu}n^{\nu}+n^{\mu}k^{\nu})\biggr]   \\
   {[}D^{-1}_{(TAG)}{]}^{\mu \nu}(k)&=&
      -k^2\biggl(g^{\mu \nu}-\frac{k^{\mu}k^{\nu}}{k^2}\biggr) -
     \frac{1}{\xi_A}n^{\mu}n^{\nu}
\eea
where $n^{\mu}=(1,0,0,0)$. 
It is noted that the gauge parameter $\xi_A$ in TAG has a dimension of 
${\rm mass}^{-2}$.
They satisfy the relations in Eq.(\ref{Identity}):
\bea
       D^{(TAG)}_{\alpha \mu}(k)[D^{-1}_{(TAG)}]^{\mu \beta}(k)&=&
    D^{(TAG)}_{\alpha \mu}(k)[-k^2d^{\mu \beta}] + 
       k_{\alpha}\biggl(\frac{n^{\beta}}{k_0}-\frac{k^{\beta}}{k^2}\biggr) 
  \nonumber  \\
       D^{-1}_{(TAG)\alpha \mu }(k)D^{\mu \beta}_{(TAG)}(k) 
          &=& [-k^2 d_{\alpha \mu}]D^{\mu \beta}_{(TAG)}(k)+ 
         \biggl(\frac{n_{\alpha}}{k_0}-\frac{k_{\alpha}}{k^2}\biggr)
         k^{\beta}
\eea
The three-gluon vertex is given by 
$\Gamma^{abc}_{\lambda \mu \nu } (p, k, q)$ in Eq.(\ref{VertexFG}).

In the limit $\xi_A=0$ the inverse of the gluon propagator does not exist.
So in the framework of PT we must work with a non-zero $\xi_A$. Then 
$\xi_A$-dependent terms appear from both the one-loop gluon 
self-energy and the corresponding pinch contributions. But again they 
cancel each other~\cite{Passera}. So we only concern about 
the $\xi_A$-independent parts of the one-loop gluon self-energy and 
pinch contributions. It is noted that although the ghost field 
decouples in the limit $\xi_A=0$, for a non-zero $\xi_A$ the ghost should 
be taken into account and at one-loop level it contributes to 
the $\xi_A$-independent part of $\Pi_{00}^{(TAG)}$~\cite{HKT}\cite{Passera}. 

Using the TAG propagator with $\xi_A=0$, 
the $\xi_A$-independent part of $\Pi_{ii}^{(TAG)}$ at one-loop was 
calculated in Ref.\cite{HKT}.  
The static limit of $\Pi_{ii}^{(TAG)}$ for $\kappa<<T$ was given in  
Eq.(4.44) of Ref.~\cite{HKT} as   
\be
  \Pi_{ii}^{(TAG)}(k_0=0,\kappa) \approx N g^2 \kappa T\frac{5}{8} +
     {\cal O} (\kappa^2)~,
\ee
from which we obtain 
\be
     \Pi_{T}^{(TAG)}(T,\kappa)=\frac{1}{2}\Pi_{ii}^{(TAG)}(0,\kappa) 
              \approx  Ng^{2}{\kappa}T\frac{5}{16}  + {\cal O} (\kappa^2).
\label{TAG}
\ee
In Ref.\cite{Sasaki} the pinch contribution to  $\Pi_{T}(T,\kappa)$
in TAG was calculated and the result was for $\kappa<<T$ 
\be
   \Pi_{T}^{P(TAG)}(T,\kappa) \approx 
        Ng^{2}{\kappa}T\frac{1}{8}+ {\cal O} (\kappa^2).
\label{PTAG}
\ee
Again the sum of $\Pi_{T}^{(TAG)}$ and $\Pi_{T}^{P(TAG)}$ coincides with 
$\Pi_{T}$ in Eq.(\ref{InvPiApprox}) and yields the same $\beta_T$ in 
Eq.(\ref{InvBeta}).  

In the evaluation of $\Pi_{ii}^{(TAG)}(k_0=0,\kappa)$ and 
$\Pi_{T}^{P(TAG)}(T,\kappa)$, there 
appear   $\vec {\bf k}^2 / \vec {\bf p}^2$ 
singularities  at the lower limit of the integration, 
due to the $1/{p_0^2}$ and $1/{q_0^2}$ terms coming from the 
TAG propagator. These singularities were  
circumvented~\cite{HKT}\cite{Sasaki} by the principal value 
prescription~\cite{KUMMER}. 
In the present paper we show instead that when the pinch contributions 
are added to $\Pi_{T}^{(TAG)}(T,\kappa)$ before the 
$p(=\vert \vec {\bf p} \vert)$-integration, 
these $\vec {\bf k}^2 / \vec {\bf p}^2$ singularities cancel and 
the limit $p \rightarrow 0$ becomes regular, so that  
we can evaluate the sum of $\Pi_T^{(TAG)}$ and $\Pi_T^{P(TAG)}$ 
without recourse to the principal value prescription. 
In fact Eq.(4.43) for $\Pi_{ii}^{(TAG)}(0,\kappa)$ in Ref.\cite{HKT} 
will be  rewritten as 
\bea
 \Pi_{ii}^{(TAG)}(0,\kappa)&=& \frac{Ng^2}{2\pi^2}\int_{0}^{\infty} dp~p~n(p)
     \nonumber \\ 
   & & \times \Biggl[-2+\frac{\kappa^2}{p^2}+
   \frac{\kappa^4}{4p^4}+\biggl(\frac{2p}{\kappa}+\frac{5\kappa}{2p}-
   \frac{\kappa^3}{2p^3}-\frac{\kappa^5}{16p^5}\biggr)
       ~{\rm ln} 
        \Big\vert \frac{2p + \kappa}{2p - \kappa} \Big\vert 
      \Biggr]  \nonumber  \\
\eea
if we do not apply the principal value prescription. We see that 
the integrand (the terms in $[\cdots]$) will behave as $-4\kappa^2/3p^2$ for 
small $p$.

Since the TAG propagator and its inverse satisfy the relation of 
Eq.(\ref{Identity}), we can follow the same procedure as before and we 
obtain for the $\xi_A$-independent part of the 
pinch contribution to the gluon self-energy in TAG, 
\bea
  \Pi^{\mu \nu}_{P(TAG)} &=& N g^2  k^2 d^{\mu \nu} \int dp 
     \frac{1}{p^2 q^2 p_0^2 } (k^2 +2p^2-  q^2 -
 4 {\vec {\bf k}}\cdot {\vec {\bf p}})     \nonumber \\
    &+&\frac{N}{2}g^2  k^2 d^{\mu \alpha} d^{\nu \beta}\int dp 
\frac{1}{p^2 q^2 p_0^2 q_0^2}
\biggl\{p_{\alpha}p_{\beta}(k^2- 4q_0^2+2{\vec {\bf k}}^2 )
           \nonumber \\ 
& &\quad  +(p_{\alpha}n_{\beta}+n_{\alpha}p_{\beta})
      \bigl[-p^2q_0+q^2p_0-2{\vec {\bf p}}\cdot {\vec {\bf q}}(p_0-q_0)\bigr]
     +n_{\alpha}n_{\beta} 4p_0q_0 (pq) \biggr\}  \nonumber \\
\eea
Then in the static limit, $\Pi_{T}^{P(TAG)}$ is expressed as 
\bea
     \Pi_{T}^{P(TAG)}(T,\kappa)&=& - N g^2 {\kappa}^2 
   \int dp \biggl\{ 
   \frac{{\vec {\bf k}}^2 + 4{\vec {\bf k}} \cdot {\vec {\bf p}}}
        {p^2 q^2 p_0^2} + \frac{1}{p^2  p_0^2} -  \frac{2}{q^2 p_0^2} \biggr\} 
       \nonumber   \\
     & & - \frac{N}{4}g^2 {\kappa}^2 \int dp 
      \biggl[ \vec {\bf p}^2-\frac{(\vec {\bf k}\cdot \vec {\bf p})^2}
           {\vec {\bf k}^2} \biggr]
    \biggl\{ \frac{\vec {\bf k}^2}{p^2 q^2 p_0^2 q_0^2} - 
        \frac{4}{p^2 q^2 p_0^2}  \biggr\}.
\label{PiPTAG}
\eea
where the terms proportional to $(p_{\alpha}n_{\beta}+n_{\alpha}p_{\beta})$ 
and $n_{\alpha}n_{\beta}$ in $\Pi^{\mu \nu}_{P(TAG)}$ do not contribute 
to $\Pi_{T}^{P(TAG)}$. 
After the $p_0$-summation and the angular integration, 
$\Pi_{T}^{P(TAG)}(T,\kappa)$ is rewritten as 
\bea
 \Pi_{T}^{P(TAG)}(0,\kappa)&=& \frac{Ng^2}{4\pi^2}\int_{0}^{\infty} dp~p~n(p)
     \nonumber \\ 
   & & \times \Biggl[-\frac{\kappa^2}{p^2}-
   \frac{\kappa^4}{4p^4}+\biggl(\frac{\kappa}{p}
   +\frac{\kappa^3}{2p^3}+\frac{\kappa^5}{16p^5}\biggr)
       ~{\rm ln} 
        \Big\vert \frac{2p + \kappa}{2p - \kappa} \Big\vert 
      \Biggr]~,  \nonumber  \\
\eea
where we have used formulas given in Appendix B.
Note that the integrand behaves as $4\kappa^2/3p^2$ for small $p$. 
When $\Pi_{T}^{(TAG)}$ and $ \Pi_{T}^{P(TAG)}$ are combined 
(remember $\Pi_{T}^{(TAG)}=\frac{1}{2}\Pi_{ii}^{(TAG)}(0,\kappa)$), 
the $\kappa^2/p^2$ singularities cancel and the integrand becomes regular as 
$p \rightarrow 0$.  Indeed we find the sum of 
$\Pi_{T}^{(TAG)}$ and $ \Pi_{T}^{P(TAG)}$ 
coincides with $\Pi_{T}$ in Eq.(\ref{InvPi}).

\bigskip
\section{Summary and Discussion}

The calculation of 
the thermal $\beta$ function $\beta_T$ was performed in four
different gauges, that is, in BFM with an arbitrary gauge parameter, 
in FG, in CG, and 
in TAG. When the pinch contributions were taken care of, the same result  
$\beta_T=g^3 N\frac{7}{32}\frac{T}{\kappa}$ 
was obtained at one-loop order in all four cases. 

However, this is not the end of the story. Elmfors and Kobes 
pointed out~\cite{EK} that the leading contribution to $\beta_T$, 
which gives a term $T/\kappa$,  does not come from 
the hard part of the loop integral, responsible  for a $T^2/\kappa^2$ term, 
but from soft loop integral. Hence they emphasized that it is not consistent 
to stop the calculation at one-loop order for soft internal momenta and that  
the resummed propagator and vertices~\cite{BP} must be used to get the 
complete leading contribution. The need for resummation is urged also by 
the following observation: The fact that we have obtained  
the same $\beta_T$ at one-loop level in four different gauges
implies that the effective gluon self-energy 
$\widehat{\Pi}^{\mu \nu}$ in Eq.(\ref{InvSelf}), constructed in BFM or in FG  
with recourse to PT, is gauge-fixing independent and universal. 
Provided that we use $\widehat{\Pi}^{\mu \nu}$ for calculation of the 
gluon damping rate $\gamma$ at zero momentum, we would obtain 
$\gamma=-11Ng^2T/(24\pi)$~\cite{HZ}, a negative damping rate, which 
is not acceptable today~\cite{BP}\cite{KKR}.

Since the corrections to the bare 
propagator and vertices, which come from the hard thermal loops, 
are gauge-independent and satisfy simple Ward identities~\cite{BP}, 
it is well-expected that we will obtain the gauge-independent thermal 
$\beta$ function even when we use the resummed propagator and vertices 
in the framework of PT. Study along this direction is under way.

\bigskip
\bigskip

I would like to thank Professor A. Sirlin for the hospitality extended to me 
at New York University where this work was completed and for 
helpful discussions. I am also  
grateful to Professor D. Zwansiger, M. Schaden, J. Papavassiliou, 
K. Philippides and M. Passera  for critical comments and discussions.
This work is supported by Yokohama National University 
Foundation. 

\newpage
\appendix

\section{Pinch contributions to gluon self-energy}

In this Appendix we give the one-loop pinch contributions to the gluon 
self-energy from the vertex diagrams of the 
first kind ($V_1$), 
the vertex diagrams of the second kind ($V_2$), and 
box diagrams ($Box$), separately, calculated in four different gauges: 
(i) The background field method with an arbitrary $\xi_Q$; 
(ii) The Feynman gauge; (iii) The Coulomb gauge; (iv) The 
temporal axial gauge. The results in the cases of the 
background field method~\cite{Papavass} and the Feynman gauge~\cite{rCP} are 
already known, but they are listed again for completeness. 
The expressions are in the imaginary time formalism and thus
\be 
     \int dp =\int \frac{d^3 p}{(2\pi)^3} ~T \sum_{n}
\ee 
where the summation goes over $p_0=2\pi inT$. They are transformed 
into the ones in the Minkowski space by the replacement 
$\int dp \Rightarrow -i\int d^4 p/(2\pi)^4$.

\subsection{\it The background field method with an arbitrary $\xi_Q$}

\bea
 \Pi^{\mu \nu (V_1)}_{P(BFM)}&=& N g^2 (1-\xi_Q) k^2 d^{\mu \nu} \int dp 
                 \frac{1}{p^4}   \\
\ \nonumber \\
  \Pi^{\mu \nu (V_2)}_{P(BFM)} &=& N g^2 (1-\xi_Q) k^2 d^{\mu \nu} \int dp 
     \biggl[-\frac{1}{p^2 q^2}-\frac{4kp}{p^4 q^2} \biggr]  \nonumber \\
& & +\frac{N}{2}g^2 (1-\xi_Q)^2 k^4 d^{\mu \alpha}d^{\nu \beta}\int dp 
  \frac{-2p_{\alpha}p_{\beta}}{p^4 q^4}   \\
\   \nonumber \\
  \Pi^{\mu \nu (Box)}_{P(BFM)} &=& N g^2 (1-\xi_Q) k^4 d^{\mu \nu} \int dp 
     \frac{-1}{p^4 q^2} \nonumber \\
& & +\frac{N}{2}g^2 (1-\xi_Q)^2 k^4 d^{\mu \alpha}d^{\nu \beta}\int dp 
  \frac{p_{\alpha}p_{\beta}}{p^4 q^4} 
\eea

\subsection{\it The Feynman gauge}

There are no contributions from the box and the vertex diagrams of the first 
kind. Thus we have 
$ \Pi^{\mu \nu (V_1)}_{P(FG)}= \Pi^{\mu \nu (Box)}_{P(FG)}=0$.
The only contribution to the pinch part comes from the vertex diagrams of the 
second:
\be
 \Pi^{\mu \nu (V_2)}_{P(FG)}=2 N g^2 k^2 d^{\mu \nu} \int dp 
     \frac{1}{p^2 q^2}.
\ee

\subsection{\it The Coulomb gauge}

Only the $\xi_C$-independent parts of the pinch contributions are listed.
\bea
 \Pi^{\mu \nu (V_1)}_{P(CG)}&=&N g^2  k^2 d^{\mu \nu} \int dp 
                 \frac{-1}{p^2{\vec {\bf p}}^2}   \\
\  \nonumber \\
  \Pi^{\mu \nu (V_2)}_{P(CG)} &=& N g^2  k^2 d^{\mu \nu} \int dp 
     \frac{-4 {\vec {\bf k}}\cdot {\vec {\bf p}}}
          {p^2 q^2 {\vec {\bf p}}^2}  \nonumber \\
& &+ N g^2  k^2 d^{\mu \alpha} d^{\nu \beta} \int dp 
     \frac{1}{p^2 q^2 {\vec {\bf p}}^2 {\vec {\bf q}}^2}   
    \biggl\{ p_{\alpha}p_{\beta} ({\vec {\bf k}}^2 - 2 {\vec {\bf q}}^2 )  
          + n_{\alpha}n_{\beta} p_0 q_0 (k^2 + 2 pq)   
   \nonumber  \\
& &\qquad + (p_{\alpha}n_{\beta}+ n_{\alpha}p_{\beta})\frac{1}{2}
  \bigl[p_0 p^2-q_0 q^2+ 2(p_0-q_0)(p_0 q_0 -  
         2 {\vec {\bf p}}\cdot {\vec {\bf q}} ) \bigr] \biggl\}
                \nonumber  \\
& &+ \frac{N}{2} g^2 \Biggl[ d^{\mu \alpha} \int dp 
  \biggl\{ p_{\alpha}k^{\nu} \biggl[ \frac{1}{q^2 \vec {\bf p}^2} - 
      \frac{1}{p^2 \vec {\bf q}^2} + 
   \bigl( \frac{1}{q^2}-\frac{1}{p^2} \bigr) 
   \frac{{\vec {\bf p}}\cdot {\vec {\bf q}} }
   {\vec {\bf p}^2 \vec {\bf q}^2}  \biggr]  \nonumber \\ 
& & \qquad \qquad + n_{\alpha}k^{\nu}\biggl[ 
  - \frac{q_0}{p^2 {\vec {\bf q}}^2} -\frac{p_0}{q^2 {\vec {\bf p}}^2}
     +   \bigl(\frac{q_0}{q^2}+\frac{p_0}{p^2} \bigr) 
      \frac{{\vec {\bf p}}\cdot {\vec {\bf q}}}
       {  {\vec {\bf p}}^2 {\vec {\bf q}}^2}  \biggr] \biggr\}
         + (\mu \leftrightarrow \nu) \Biggr]    \\
\   \nonumber \\
 \Pi^{\mu \nu (Box)}_{P(CG)} &=& N g^2  k^4 d^{\mu \nu} \int dp
     \frac{1}{p^2 q^2 {\vec {\bf p}}^2}  \nonumber \\   
  & & +\frac{N}{2}g^2  k^4 d^{\mu \alpha} d^{\mu \beta}\int dp 
\frac{1}{p^2 q^2 {\vec {\bf p}}^2 {\vec {\bf q}}^2}
\bigl[p_{\alpha}p_{\beta} +  (p_{\alpha} n_{\beta}+ n_{\alpha} p_{\beta})
 ( q_0 - p_0 )
  \nonumber \\
 & &\ \ \ \ \ \ \ \ \ \ \ \ \ \ \ \ \ \ \ \ \ \ \ \ \ \ \ \ \ \ \  
 \qquad \qquad          
 - 2 n_{\alpha} n_{\beta} p_0 q_0 \bigr]
\eea        

\subsection{\it The temporal axial gauge}

Only the $\xi_A$-independent parts of the pinch contributions are listed.
\bea
 \Pi^{\mu \nu (V_1)}_{P(TAG)}&=& N g^2  k^2 d^{\mu \nu} \int dp 
   \frac{-1}{p^2 p_0^2}    \\
\   \nonumber \\
  \Pi^{\mu \nu (V_2)}_{P(TAG)} &=& N g^2  k^2 d^{\mu \nu} \int dp
      \frac{1}{p^2 q^2} \biggl\{ \frac{2 p^2}{p_0^2} 
  - \frac{4{\vec {\bf k}}\cdot {\vec {\bf p}}}{p_0^2} 
              \biggr\} \nonumber \\
& &+ N g^2  k^2 d^{\mu \alpha} d^{\nu \beta} \int dp 
     \frac{1}{p^2 q^2 p_0^2 q_0^2}   
    \biggl\{ p_{\alpha}p_{\beta}   \bigl[ 
         {\vec {\bf k}}^2 - 2 q_0^2 \bigr] + 
   n_{\alpha}n_{\beta} p_0 q_0 (k^2 + 2 pq)   
   \nonumber  \\
& &\quad + (p_{\alpha}n_{\beta}+ n_{\alpha}p_{\beta})
  \bigl[ \frac{1}{2} (p_0 p^2-q_0 q^2) -p_0 ({\vec {\bf q}}^2+
  2{\vec {\bf p}}\cdot {\vec {\bf q}}) 
   +q_0 ({\vec {\bf p}}^2+ 2{\vec {\bf p}}\cdot {\vec {\bf q}})               
                \bigr] \biggl\}
                \nonumber  \\
                         \\
\    \nonumber \\
  \Pi^{\mu \nu (Box)}_{P(TAG)} &=& N  g^2  k^4 d^{\mu \nu} \int dp
     \frac{1}{p^2 q^2 p_0^2}     \nonumber \\   
  & &+\frac{N}{2}g^2  k^4 d^{\mu \alpha} d^{\nu \beta}\int dp 
\frac{p_{\alpha}p_{\beta} + 
   (p_{\alpha} n_{\beta}+ n_{\alpha} p_{\beta}) (q_0-p_0) 
  - 2 n_{\alpha} n_{\beta}p_0 q_0 }{p^2 q^2 p_0^2 q_0^2}       
\eea

\section{Thermal one-loop integrals}

We list the thermal one-loop integrals 
in the static limit $k_0=0$ which are 
used in this paper. 
We only give the matter part.
Due to the constraint $k+p+q=0$ there holds a relation 
\be
       \int dp f(p,q) = \int dp f(q,p).
\ee
It is understood that 
in the r.h.s. of the expressions below, $p\equiv \vert \vec {\bf p} \vert$,  
$\kappa\equiv \vert \vec {\bf k} \vert$ and $n(p)=1/[{\rm exp} (p/T) -1 ]$. 
\bea
   \int dp \frac{p_0^2}{p^2 q^2}&=&\frac{1}{4\pi^2 \kappa}
        \int_{0}^{\infty} dp ~p^2 ~n(p) ~{\rm ln} 
        \Big\vert \frac{2p + \kappa}{2p - \kappa} \Big\vert   \\ 
\   \nonumber \\
    \frac{1}{\vec {\bf k}^2} \int dp 
    \frac{(\vec {\bf k} \cdot \vec {\bf p})^2 }{p^2 q^2}&=&
     \frac{1}{8\pi^2 }
        \int_{0}^{\infty} dp ~p ~n(p) 
         \biggl\{ 2 + \frac{\kappa}{2p}~{\rm ln} 
   \Big\vert \frac{2p + \kappa}{2p - \kappa} \Big\vert \biggr\}  \\
\   \nonumber \\  
    \vec {\bf k}^2 \int dp \frac{1}{p^2 q^2}&=&\frac{\kappa}{4\pi^2}
        \int_{0}^{\infty} dp ~n(p) ~{\rm ln} 
        \Big\vert \frac{2p + \kappa}{2p - \kappa} \Big\vert  \\
\   \nonumber \\
    \vec {\bf k}^2 \int dp 
    \frac{\vec {\bf k} \cdot \vec {\bf p} }{p^4 q^2}&=&
     \frac{\kappa}{8\pi^2 }
        \int_{0}^{\infty} dp ~n(p) 
         \biggl\{ - \frac{4p \kappa}{(2p+\kappa)(2p-\kappa)} + 
   {\rm ln}  \Big\vert \frac{2p + \kappa}{2p - \kappa} \Big\vert \biggr\}
       \nonumber         \\
\    \\
\   \nonumber \\
    \int dp 
\frac{ \vec {\bf k}^2 \vec {\bf p}^2-
     (\vec {\bf k} \cdot \vec {\bf p})^2}{p^4 q^2}&=&
    - \frac{1}{2} \vec {\bf k}^2 \int dp \frac{1}{p^2 q^2}
\eea
\be
    \vec {\bf k}^2 \int dp 
\frac{ \vec {\bf k}^2 \vec {\bf p}^2-
     (\vec {\bf k} \cdot \vec {\bf p})^2}{p^4 q^4}=\frac{\kappa}{4\pi^2}
        \int_{0}^{\infty} dp ~p~n(p) \biggl\{ -
   \frac{2\kappa^2}{(2p+\kappa)(2p-\kappa)}+\frac{\kappa}{2p}~{\rm ln} 
        \Big\vert \frac{2p + \kappa}{2p - \kappa} \Big\vert 
      \biggr\}   \nonumber \\
\ee 
\bea
   \int dp \frac{1}{p^2}&=& - \frac{1}{2\pi^2}
        \int_{0}^{\infty} dp ~p~n(p) = -\frac{1}{12}T^2  \\
\  \nonumber \\
\int dp \frac{1}{\vec {\bf p}^2}&=& 0 \quad \quad \quad 
      {\rm (for\ \  matter\ \  part)} \\
\   \nonumber \\
\int dp \frac{p_0^2}{p^2 \vec {\bf q}^2}&=& 
     - \frac{1}{4\pi^2 \kappa} 
      \int_{0}^{\infty} dp ~p^2~n(p) ~{\rm ln} 
        \Big\vert \frac{p + \kappa}{p - \kappa} \Big\vert  
\eea
\bea
 & & \int dp \frac{1}{p^2q^2}\biggl[\vert \vec {\bf p} - \vec {\bf q} \vert^2 
    (1+c^2) + 8( \vec {\bf p}^2 +\vec {\bf q}^2 )(1-c^2) \biggr]   \nonumber \\
 & & \qquad = \frac{1}{2\pi^2} \int_{0}^{\infty} dp ~p~n(p)
     \Biggl\{ 5+\frac{5}{2}\frac{\kappa^2}{p^2}+
 \frac{(p^2-\kappa^2)^2(6p^2+\kappa^2)}{4p^5 \kappa}~{\rm ln}
   \Big\vert \frac{p + \kappa}{p - \kappa} \Big\vert  \nonumber \\
 & & \qquad \qquad \qquad  \qquad 
    +~\frac{32p^6+40p^4\kappa^2-8p^2\kappa^4-\kappa^6}
  {8p^5\kappa} ~{\rm ln} \Big\vert \frac{2p + \kappa}{2p - \kappa} \Big\vert 
  \Biggr\}  
\eea 
\bea
   \vec {\bf k}^2 \int dp
   \frac{1}{p^2 \vec {\bf q}^2}&=&-\frac{\kappa}{4\pi^2}
     \int_{0}^{\infty} dp~n(p)
   ~{\rm ln} 
        \Big\vert \frac{p + \kappa}{p - \kappa} \Big\vert  \\
\ \nonumber \\
    \vec {\bf k}^2 \int dp
   \frac{\vec {\bf k} \cdot \vec {\bf p}}{p^2 q^2 \vec {\bf p}^2}&=&
     \frac{\kappa}{4\pi^2} \int_{0}^{\infty} dp~n(p) 
      \Biggl\{ \frac{\kappa}{p} +\frac{\kappa^2-p^2}{2p^2}
       ~{\rm ln} \Big\vert \frac{p + \kappa}{p - \kappa} \Big\vert-
     \frac{\kappa^2}{2p^2} 
    ~{\rm ln} \Big\vert \frac{2p + \kappa}{2p - \kappa} \Big\vert 
    \Biggr\}   \nonumber \\
\eea
\bea
   \vec {\bf k}^2 \int dp \biggl[ 1-\frac{(\vec {\bf k} \cdot \vec {\bf p})^2}
   {\vec {\bf k}^2 \vec {\bf p}^2} \biggr] \frac{1}{p^2 q^2} &=& 
    \frac{\kappa}{4\pi^2} \int_{0}^{\infty} dp~n(p) \Biggl\{ 
       -\frac{p}{2\kappa} +\frac{\kappa}{2p}   \nonumber \\
 & & \quad 
       - \biggl[1-\frac{(\kappa^2+p^2)^2}{4\kappa^2 p^2} \biggr]
      ~{\rm ln} \Big\vert \frac{p + \kappa}{p - \kappa} \Big\vert +
    \biggl(1-\frac{\kappa^2}{4p^2} \biggr) 
     ~{\rm ln} \Big\vert \frac{2p + \kappa}{2p - \kappa} \Big\vert
    \Biggr\}   \nonumber  \\
\   \\
\    \nonumber \\
   \vec {\bf k}^4 \int dp \biggl[ 1-\frac{(\vec {\bf k} \cdot \vec {\bf p})^2}
   {\vec {\bf k}^2 \vec {\bf p}^2} \biggr]
        \frac{1}{p^2 q^2 \vec {\bf q}^2} &=& 
    \frac{\kappa}{4\pi^2} \int_{0}^{\infty} dp~n(p) \Biggl\{ 
    - \frac{\kappa}{p} \nonumber \\
 & &  
   - \biggl[1-\frac{(\kappa^2+p^2)^2}{4\kappa^2 p^2} \biggr]
  \frac{2\kappa^2}{p^2}
      ~{\rm ln} \Big\vert \frac{p + \kappa}{p - \kappa} \Big\vert +
   \biggl(1-\frac{\kappa^2}{4p^2} \biggr)\frac{\kappa^2}{p^2} 
     ~{\rm ln} \Big\vert \frac{2p + \kappa}{2p - \kappa} \Big\vert
    \Biggr\} \nonumber \\  
\eea
\bea
   \vec {\bf k}^2 \int dp \frac{\vec {\bf k}^2 + 
  4 \vec {\bf k} \cdot \vec {\bf p}}{p^2q^2p_0^2} &=&
   \frac{1}{4\pi^2}\int_{0}^{\infty} dp~p~n(p)
  \biggl( -\frac{\kappa^3}{p^3} \biggr)
    ~{\rm ln} \Big\vert \frac{2p + \kappa}{2p - \kappa} \Big\vert  \\
\  \nonumber \\
   \vec {\bf k}^2 \int dp \frac{1}{q^2p_0^2} &=&
   \frac{1}{4\pi^2}\int_{0}^{\infty} dp~p~n(p)
      \biggl( -2\frac{\kappa^2}{p^2} \biggr)  \\
\ \nonumber \\
   \vec {\bf k}^2 \int dp \frac{1}{p^2p_0^2} &=&
   \frac{1}{4\pi^2}\int_{0}^{\infty} dp~p~n(p)
      \biggl( -2\frac{\kappa^2}{p^2} \biggr)  
\eea
\bea
& & \vec {\bf k}^4 \int dp \biggl[
        \vec {\bf p}^2-\frac{(\vec {\bf k} \cdot \vec {\bf p})^2}
        {\vec {\bf k}^2} \biggr] \frac{1}{p^2q^2p_0^2q_0^2} 
    \nonumber \\
& & \qquad \qquad  = \frac{1}{4\pi^2}\int_{0}^{\infty} dp~p~n(p)
   \Biggl\{\frac{\kappa^4}{p^4} +\frac{\kappa^3(4p^2-\kappa^2)}{4p^5}
   ~{\rm ln} \Big\vert \frac{2p + \kappa}{2p - \kappa} \Big\vert 
    \Biggr\}  \\
\  \nonumber \\
& & \vec {\bf k}^2 \int dp \biggl[
        \vec {\bf p}^2-\frac{(\vec {\bf k} \cdot \vec {\bf p})^2}
        {\vec {\bf k}^2} \biggr] \frac{1}{p^2q^2p_0^2} 
  \nonumber \\
& & \qquad \qquad  = \frac{1}{4\pi^2}\int_{0}^{\infty} dp~p~n(p)
    \Biggl\{\frac{\kappa^2}{p^2} +\frac{\kappa(4p^2-\kappa^2)}{4p^3}
   ~{\rm ln} \Big\vert \frac{2p + \kappa}{2p - \kappa} \Big\vert 
    \Biggr\} 
\eea

\medskip

For the $\kappa << T$ expansion we use the 
following formulas~\cite{HKT}:
\bea
  \frac{1}{2}~{\rm ln} 
   \Big\vert \frac{1+x}{1-x} \Big\vert &=& \sum_{r=1}\frac{1}{2r-1}
         x^{2r-1} \qquad \qquad (x<1)      \\
           &=&\sum_{r=1}\frac{1}{2r-1}x^{1-2r} \qquad \qquad (x>1) \\
\  \nonumber \\
   \int_{0}^{1} dx \frac{x^s}{e^{yx}-1}&=& 
    \frac{1}{sy}+ \cdots      \\
   \int_{1}^{\infty} dx \frac{1}{x^s(e^{yx}-1)}&=& 
       \frac{1}{sy}+ \cdots     \\
          &=& \frac{1}{y}+\frac{1}{2} ~{\rm ln} y + \cdots  , 
  \qquad {\rm if}\ \ s=1
\eea


\bigskip
\newpage
\medskip
\noindent
{\large\bf Figure caption}
\medskip

\noindent
Fig.1

\noindent
The self-energy diagram for the quark-quark scattering.

\medskip

\noindent
Fig.2

\noindent
(a) The vertex diagrams of the first kind for 
the quark-quark scattering. 
(b) Their pinch contribution.

\medskip

\noindent
Fig.3

\noindent
(a) The vertex diagram of the second kind 
for the quark-quark scattering. 
(b) Its pinch contribution.

\medskip

\noindent
Fig.4

\noindent
(a) The box diagrams for the quark-quark scattering.   
(b) Their pinch contribution.

\end{document}